\documentclass[fleqn,usenatbib]{mnras}
\usepackage[T1]{fontenc}
\usepackage{ae,aecompl}
\usepackage{graphicx}	% Including figure files
\usepackage{amsmath}	% Advanced maths commands
\usepackage{amssymb}	% Extra maths symbols
\usepackage{enumerate}   % For making custom lists 
\usepackage{booktabs}   % Extra table functions
\usepackage{float}
\usepackage{newtxtext,newtxmath}

\newcommand{\unit}[1]{\ensuremath{\, \mathrm{#1}}}  % Shorthand for proper spacing with units.

\title[Error effects]{Observation uncertainty effects on the precision of interior planetary parameters}

\author[Plotnykov \& Valencia]{
Mykhaylo Plotnykov,$^{1}$
Diana Valencia$^{2,3}$
\\
$^{1}$Department of Physics, University of Toronto, Toronto, ON M5S 3H4, Canada\\
$^{2}$Department of Physical \& Environmental Sciences, University of Toronto at Scarborough, Toronto, ON M1C 1A4, Canada\\
$^{3}$Department of Astronomy \& Astrophysics, University of Toronto, Toronto, ON M5S 3H4, Canada\\
}

\date{Accepted XXX. Received YYY; in original form ZZZ}

\pubyear{2023}

\begin{document}
\label{firstpage}
\pagerange{\pageref{firstpage}--\pageref{lastpage}}
\maketitle
% Abstract of the paper
\begin{abstract}
   Determining compositions of low-mass exoplanets is essential in understanding their origins. The certainty by which masses and radius are measured affects our ability to discern planets that are rocky or volatile rich. In this study, we aim to determine sound observational strategies to avoid diminishing returns. We quantify how uncertainties in mass, radius and model assumptions propagate into errors in inferred compositions of rocky and water planets.  For a target error in a planet's iron-mass fraction or water content, we calculate the corresponding required accuracies in radius and mass. For instance, a rocky planet with a known radius error of 2$\%$ (corresponding to TESS detection best errors) demands mass precision to be at 5-11$\%$ to attain a 8 wt$\%$ precision in iron-mass fraction, regardless of mass. Similarly, a water world of equal radius precision requires 9-20$\%$ mass precision to confine the water content within a 10 wt$\%$ margin. Lighter planets are more difficult to constrain, especially water-rich versus water-poor worlds. Studying Earth as an exoplanet, {we find a $\sim \pm5$ point "error floor" in iron-mass fraction and $\sim \pm 7$ in core-mass fraction from our lack of knowledge on mineralogy.} The results presented here can quickly guide observing strategies to maximize insights into small exoplanet compositions while avoiding over-observing.
\end{abstract} 

%%%%%%%%%%%%%%%%%%%%%%%%%%%%%%%%%%%%%%%%%%%%%%%%%%

%%%%%%%%%%%%%%%%% BODY OF PAPER %%%%%%%%%%%%%%%%%%

\section{Introduction}
    Over the years, improvements in instrument technology and data analysis have lead to major advances in exoplanet sciences. 
    Low-mass exoplanets, both super-Earths and mini-Neptunes, are now routinely detected and their radius and mass are commonly being measured despite their small sizes. 
    This observational progress has lead to an increase in sample size that lends itself to population studies, albeit with challenges arising from the heterogeneity of the data \citep{Teske2021}. 
    Given the overlap in mass-radius space for super-Earths and mini-Neptunes, the first task for characterizing these planets is to obtain their bulk composition, which entails inferring their bulk chemical inventory and determining whether a planet is rocky, ocean or gaseous.  
    Our ability to make these inferences resides in accurate observations of both planetary masses and radii (\citealt{Valencia2007,Rogers2010,Zeng2013,Dorn2015,Plotnykov2020}).
    
    Fortunately, the precision in radius measurements has improved over the years with each new mission/instrument launched, from ground based telescopes, to Kepler, to recently TESS. 
    Nowadays, the TESS mission is consistently yielding a precision of $\sim 5\%$ in radius for low-mass exoplanets, with some observations even having errors of $\sim 1\%$ \citep{Bonfanti2021}. 
    The newer generation of telescopes such as JWST, Ariel, PLATO and others will continue to provide better data (\citealt{Gardner2006,Rauer2014,Tinetti2018}).
    However, a strategy that improves the precision of one of the parameters, such as radius, without improving the other (i.e. mass) will not guarantee a better estimate of the planetary composition. 
    Thus, a sound strategy that considers in tandem the effect of mass and radius precision on the compositional inference is more effective and avoids wasting observational time.
    At the moment, the limitation stems from the lack of precise mass estimates with $\sim 20\%$ average error for low-mass exoplanets (up to 10 $M_\oplus$ from NASA exoplanet archive\footnote{ \url{https://exoplanetarchive.ipac.caltech.edu}}).
    The reason is that improvements in mass measurements require expensive observational campaigns.
    For example, 55~Cnc~e was studied over a decade with numerous ground base telescopes leading to more than 1500 radial velocity measurements, to achieve a precision of $<4\%$ in mass \citep{Bourrier2018}. 
    These lengthy mass follow up campaigns are expected to be more manageable with the newer spectrographs that are in operation or coming online in the next few years (MAROON-X, ELT/METIS, SPIRou, HARPS, etc.). 
    The question becomes, how well do we need to estimate mass and radius to obtain useful compositional information? 
    We aim to answer this question.
    
    Establishing accurately the chemical composition of these low-mass exoplanets opens a unique window to understanding how they form. 
    Contrary to exo-Jupiters that clearly acquire large quantities of H-He, a successful formation theory for low-mass exoplanets is nuanced. 
    It has to explain why some planets are bare rocks and others have water and/or H-He in the observed quantities while having similar masses and radii. 
    We proposed a new and simple line of inquiry that compares the chemical ratios of rock forming elements -- or refractory ratios -- of rocky exoplanets and stars, as to reveal formation pathways \citep{Plotnykov2020}. 
    This has been expanded by other works \citep{ Adibekyan2021, Schulze2021, Wang2022}.
    The idea arises from the long upheld notion that the first solids condensing out of the disk nebula to acquire the disk chemical signature (\citealt{Larimer1967,Grossman1972}).
    These first solids eventually become the rocky portion of planets and without major chemical reprocessing during formation, the resulting planets are expected to have the same refractory ratios (Fe/Mg, Fe/Si, etc.) as their host star. 
    Deviations between the stellar and planetary refractory ratios would be indicative of chemical reprocessing and could possibly constrain formation pathways (\citealt{Asphaug2006,Carter2015,Scora2020,Clement2021}).
    Thus, comparing the Fe/Si or Fe/Mg ratios between planets and stars is a first step into understanding the origin and evolution of these planets.  
    We showed that rocky super-Earths display a wider chemical range than that of stars in a population sense \citep{Plotnykov2020}. 
    One-to-one comparisons (\citealt{Plotnykov2020,Adibekyan2021,Schulze2021,Wang2022}) between host star and super-Earths showed that errors in chemical inventory made it difficult to draw clear conclusions.
    To deepen this line of inquiry and determine whether or not rocky super-Earths conserve their primordial composition during formation (e.g. that of the protostellar disk), it requires measuring both the stellar and planetary chemical abundances with enough precision and with unbiased data, to draw meaningful conclusions.
    
    In this regard, rocky planets are incredibly useful in understanding planetary formation, because their refractory inventory (ie. the amount of core-mass fraction) is almost uniquely determined by mass and radius data \citep{Plotnykov2020}.
    Any other planet suffers from the problem of degeneracy in composition, meaning additional parameters are required to constrain the interior. 
    This is the case for water/hycean planets, as they have three to four compositional unknowns to determine (core and mantle fractions, amount of water/ice,  and perhaps even H/He) with only two data points (mass and radius). 
    However, it is important to determine which planets have water and their amount.
    \citet{Luque2022} has suggested from observations that planets around M Dwarfs can be divided into two groups: planets that are purely rocky with composition similar to Earth and water planets that have 50\% H$_2$O content (or water mass fraction -- wmf). 
    These M dwarf planets are common observational targets owing to the fact that some of them could have habitable conditions (e.g. TRAPPIST-1, Kepler-62, GJ~9827, TOI~1452, etc.) and thus, it is important to determine the best strategy to constrain their composition as well.
    
    {In this study, we focus on determining the effect of uncertainties in planetary radius and mass, as well as mineralogy modeling, on compositional constraints. 
    We first explore the accuracy of internal structure models in general, by treating Earth as an exoplanet. 
    In this way, we can determine the intrinsic error in retrieving refractory ratios, core mass fraction (cmf) and iron-mass fraction (Fe-mf) values by benchmarking it to Earth while making the least number of assumptions. 
    Next, we consider two compositional cases: pure rocky planets and water worlds. }
    Within each case we consider two compositions and find how mass-radius errors affect inference in composition.
    We term a 'water planet' one where there is enough water to affect the structure of the planet (wmf$\ge 0.5$ wt$\%$).
    Earth's water content (at most wmf$=0.1$ wt$\%$ \citep{Bercovici2003}) has a negligible effect on its radius and thus is considered a rocky planet.
    For purely rocky planets we simulate error effects for Earth-like (cmf$=33$ wt$\%$) and Mercury-like (cmf$=74$ wt$\%$) planets.
    For water worlds we consider water-poor (wmf$=10$ wt$\%$) and water-rich (wmf$=50$ wt$\%$) planets,  partially justified by our knowledge of the water content of the icy moons of Jupiter (i.e. $50\%$ by mass \citet{Hussmann2015}).
        
    Our goals are to: (1) quantify realistic expectations on compositional uncertainties and (2) provide simple tools for observers to determine optimal observational strategies when the purpose is to constrain the iron-mass and core-mass fraction or water-mass fraction values. 

\section{Model $\&$ Method}
\subsection{Simple parameterized interior model}
\label{model:simple}    
    The observation precision of low mass exoplanets and the accuracy of interior composition inference are closely related, as small changes in mass or radius measurements can lead to significantly different chemical inventories.
    To give an idea of the importance of radius-mass precision, we use a simple analytical treatment.
    For rocky planets, we consider two chemical reservoirs, a dense core and a lighter mantle. 
    Changes in cmf ($\Delta \mathrm{cmf}$) can be expressed as following:
    \begin{equation}
    \label{eq:delta_cmf}
        \Delta \mathrm{cmf}  \approx \frac{\langle \rho_c \rangle/\rho_{bulk}}{\langle \rho_c \rangle/\langle \rho_m \rangle-1}\left[\frac{\Delta M}{M} -\frac{3\Delta R}{R}\right]
    \end{equation}
    where $\rho_{bulk}$ is the bulk density ($M/\frac{4}{3}\pi R^3$) and $\langle \rho_c \rangle$ and $\langle \rho_m \rangle$ are the average core and mantle densities, respectively. 
    The $\frac{\Delta R}{R}$, $\frac{\Delta M}{M}$ are the fractional changes in radius-mass respectively. 
    Using this formula, we can propagate uncertainties in mass-radius to find the resulting error in cmf ($\sigma_\mathrm{cmf}$):
    \begin{equation}
    \label{eq:error_cmf}
        \sigma_\mathrm{cmf}  \approx \left[\frac{\langle \rho_c \rangle/\langle \rho_m \rangle}{\langle \rho_c \rangle/\langle \rho_m \rangle-1} - \mathrm{cmf}\right] \sqrt{\left(\frac{\sigma_{M}}{M}\right)^2+9\left(\frac{\sigma_{R}}{R}\right)^2} 
    \end{equation}
    where $\sigma_{M}/M$ and $\sigma_{R}/R$ are the percentage uncertainties in mass-radius. 
    In this paper we use fractions for calculations and report the final values in percentages. 
    Consequently, the cmf errors are expressed in weight percent (i.e. a planet with cmf$=33\pm10$ wt$\%$ is equivalent to cmf$=0.33\pm0.10$).
    {In this simple treatment, the cmf is equivalent to the Fe-mf, but in general, for planets, these two measurements differ.
    Fe can be reduced in the core owing to presence of light alloys or enhanced in the mantle, depending on the degree of differentiation.}
    
    We apply a similar treatment to water worlds and find a simple-uncertainties propagation function for wmf error. 
    For this case we additionally assume that the water layer (on top of mantle) has constant average density of $\langle \rho_w \rangle$.
    Finally, the error in wmf has both dependence on mass, radius and cmf uncertainties:
    \begin{equation}
    \label{eq:error_wmf}
            \sigma_\mathrm{wmf}  \approx \sqrt{    A^2 \left[\left(\frac{\sigma_{M}}{M}\right)^2+9\left(\frac{\sigma_{R}}{R}\right)^2\right] + B^2\sigma_\mathrm{cmf}^2}
    \end{equation}
    where we define A and B as follow:
    \[
        A = \mathrm{wmf} + \frac{1+\mathrm{cmf}(\langle \rho_m \rangle/\langle \rho_c \rangle-1)}{\langle \rho_m \rangle/\langle \rho_w \rangle-1}, \ \ \ B = \frac{\langle \rho_m \rangle/\langle \rho_c \rangle-1}{1-\langle \rho_m \rangle/\langle \rho_w \rangle} \\
    \]
    We further redefine cmf to be $\mathrm{cmf} = \frac{\mathrm{rcmf}(1-\mathrm{wmf})}{\mathrm{rcmf}+1}$ where rcmf is the core to mantle mass ratio ($M_{core}/M_{mantle}$).  
    This latter parameter would be the same for planets having similar refractory ratios (e.g. Fe/Mg, Fe/Si).
    
    This very simple treatment indicates that errors in cmf/wmf are linked to actual values of the planet's cmf/wmf (through a linear dependence). 
    This means that errors will propagate slightly differently for Earth-like vs Mercury-like or water-poor vs. water-rich planets. 
    Mercury-like planets will have lower $\sigma_\mathrm{cmf}$ compared to Earth-like planets given the same relative mass-radius uncertainties.
    Meanwhile, for water worlds our simple prescription shows that  $\sigma_\mathrm{wmf}$ increases with wmf, regardless of composition of the rocky part (rcmf).
    Unsurprisingly, the consistent feature for both rocky and water planets is radii precision, where the cmf/wmf error is $\sim3$ times more sensitive to $\frac{\sigma_{R}}{R}$ compared to other parameters.
    However, this simple model assumes that the average densities of core and mantle are constant and do not change with planetary mass or composition, ignoring compression effects.
    That is, the core,  mantle and water densities in equation \ref{eq:error_cmf}, \ref{eq:error_wmf} actually depend on mass (via pressure) and for vapour planets it can also depend on temperature, so that the effect of errors in mass and radius on our inferences of cmf and wmf deviate from these simple recipes.
    To investigate how these errors depend on composition and mass, we use a sophisticated numerical interior model described in Sec. \ref{model:interior}.

\subsection{Interior model}
\label{model:interior}
    To quantify the effects of observational uncertainty on inference of planetary composition, we use the interior structure model \texttt{SuperEarth} developed by \citealt{Valencia2006, Valencia2007} and revamped by \citealt{Plotnykov2020}. 
    This model divides the planet into three main parts: core, mantle and water/ices.
    The structure equations for density, pressure, gravity, mass and temperature while assuming hydrostatic equilibrium throughout are simultaneously solved in each layer of the planet.
    The resulting structure is therefore, uniquely determined by the input parameters that include planetary mass, surface values for temperature, pressure and chemical composition (with corresponding equations of state) for the planet.
    In this work, we set pressure and temperature at 1bar and 300K at the surface respectively. 
    The exact value for the surface conditions affect interior characteristics such as the thickness of the ocean layer, but do not affect the bulk water content provided that water can condense into liquid/ice form on the planet. 
    For this study, we do not consider surface temperatures that allow for water vapor. 
    As these planets have extended atmospheres that influence considerably the total radius (\citealt{Turbet2019, Pierrehumber2023}), our results can be considered upper values for all water worlds.
  
    Our input to the interior structure model is the total mass and chemical inventory that we define by setting the mineralogy in the mantle, core and water layer as well as the cmf and wmf of the planet. 
    The output is the interior structure profiles for density, pressure, mass, gravity and temperature, including the total radius of the planet. 
    Therefore, we can represent this model as some function $f$, that takes mass and chemical composition ($\chi$) and outputs radius of the planet:  $R = f(M,\chi)$.
    To speed up our calculations we construct a grid of models by sampling in masses between 0.3-20 M$_\oplus$ and compositions that span from 0-1 of both cmf and wmf. 
    Interpolating within this grid causes a slight loss ($\lesssim 1\%$) in accuracy when determining the interior structure. 
    Note, that solving simultaneously for both cmf and wmf is impossible as the problem is degenerate, therefore for water planets, we fix the core to mantle ratio during our analysis. 
    This corresponds to effectively fixing the refractory ratios Fe/Si and Fe/Mg ratios.

    The model we use is a forward one, that calculates radius for a given mass and composition. 
    Given that we need to obtain the composition given mass and radius, this involves inverting the problem. We do so by sampling the cmf/wmf space and determining the parameter space that best fits planetary mass and radius data.
    Commonly, to solve this type of problem, a Markov Chain Monte Carlo (MCMC) sampler is coupled to an interior model of choice (\citealt{Santos2015,Dorn2017,Plotnykov2020,Agol2021,Acuna2021}), but this technique is computationally expensive. 
    To speed up retrievals, \citet{Baumeister2023} proposed a different approach using mixture density networks.
    For our purposes we use a Monte Carlo (MC) scheme because in this way we can sample parameter space thoroughly without a concern of losing computational efficiency as we are using a grid-based function. 
    
    For our simulations, we consider a mass-radius pair to which we add an uncertainty value to account for observational error assuming a Gaussian distribution.
    We keep mean values of the mass-radius fixed and vary the uncertainty around them.
    This gives us a base model for comparing different composition configurations and the effects of uncertainty on them. 
    We draw 50,000 mass-radius pairs from this distribution and determine the probability distribution of both cmf or wmf. 
    We also check if the drawn pairs are physically possible and if not, we re-draw and find a new mass-radius pair.
    In the case for purely rocky planets the physical constraint is that for a given mass the radius cannot be too compact (density higher than pure iron) or too puffy (above the rocky threshold radius).
    Same principle is applied for water planets, though there is a higher radius freedom as more mass-radius combinations are possible.
    We make our code available at \href{https://github.com/mplotnyko/exopie}{https://github.com/mplotnyko/exopie}, that can easily be used to find interior composition distribution given mass-radius.

\subsection{Rocky and Water Planets}
\label{model:MC}
    We categorize planets into two types: rocky and water planets. 
    \begin{enumerate}[I]
        \item \textbf{Rocky Planets}: {have mantle composed of magnesium (Mg), silicon (Si), iron (Fe), oxygen (O) and core primarily composed of Fe alloy with some Si, nickel (Ni) and trace elements. We assume a range of possible chemical inventories ($\chi$)  instead of fixing it to Earth's. We vary the two chemical parameters that have the most influence in the structure and radius of a planet: (1) the amount of alloy in the core -- that for the purposes of comparing planet-star compositions we take solely as Si ($\mathrm{xSi}$)  and (2) the amount of iron in the mantle, commonly expressed as the mantle magnesium number ($\mathrm{Mg}\# = \frac{\mathrm{Mg}}{\mathrm{Mg}+\mathrm{Fe}} = 1-\mathrm{xFe}$)\footnote{This equality is true only if Mg species present and Fe is replacing the Mg, thus the molar fraction of Fe+Mg adds up to 1.}, which is a measure of mantle-core differentiation. We consider $\mathrm{xSi}$ in the core to be between $0-20\%$ by mol and the rest is $68-88\%$ Fe, $10\%$ Ni and $2\%$ trace elements by mol (to account for Co, Cr, Mn and P), thus making an Fe-Ni-Si alloy. The mantle has a variable $\mathrm{xFe}$ between $0-20\%$ by mol ($\mathrm{Mg}\#=1-0.8$). We fix the amount of magnesiowustite ($\mathrm{xWu}$) in the lower mantle to $20\%$, see appendix \ref{appendix_A} for the reasoning behind these choices.} Finally, we determine error effects for an Earth-like planet with cmf$=33$ wt$\%$ and Mercury-like with cmf$=74$ wt$\%$ compositions.
        \item \textbf{Water Planets}: are differentiated with the water layer above a rocky interior. We consider the water layer to be pure H$_2$O in condensed state (liquid/ice). We examine, as well,  three scenarios for the rocky composition: (A)  fixed values for Fe/Mg based on the stellar mean values (fixed model) (B) adopting priors for the Fe/Mg based on stellar measurements (stellar prior), (C) no assumptions (uniform/unconstrained prior). We determine error effects for water-poor (wmf$=10$ wt$\%$) and water-rich compositions (wmf$=50$ wt$\%$).
    \end{enumerate}

    We simulate planets with a mean mass of 5 $M_\oplus$ and corresponding radius for a given composition. 
    We subsequently add radius and mass uncertainties and limit them to $\sigma_R/R = 0.5-10\%$ in radius and to $\sigma_M/M =  0.5-30\%$ in mass.
    We obtain all possible Fe-mf/wmf that fit the data. 
    {We present the results for the rocky exoplanets in terms of the Fe-mf, as it is the most robust metric when making comparisons with their host star.}
    We report the results in terms of 1-standard deviation ($16^{\mathrm{th}}$ and $84^{\mathrm{th}}$ percentile) values from the derived distributions.
    In the cases where the large mass-radius uncertainties result in a skewed posterior distribution owing to the fact that there are physical constraints (cmf, wmf $\in$ [0,1]) that cause the probability to be asymmetrical, {we report the average of the upper and lower bounds. This is equivalent to the half interquartile range (IQR/2 = $(\sigma_+ + \sigma_-)/2$).}

\section{{Earth As An Exoplanet}}
\label{Earth}
    To understand the sources of error in our modeling approach we first look at the Earth as an exoplanet.
    We prioritize reproducing Earth's cmf, given the fact that there is abundant seismological data that precisely constraints the density profile and location of the core-mantle boundary. This results in an accurate cmf estimate of $32.5 \pm 0.3$ wt$\%$ \citep{Wang2018}. 
    In contrast, the total elemental abundances (e.g. Fe, Mg, Si, etc.) for Earth are less constrained, owing to the fact that the exact composition of Earth's reservoirs are unknown and thus are dependant on chemical assumptions. 
    For example, the total iron budget changes depending on the amount of alloy in the core mostly (\citealt{Fischer2015, Zhang2018, Umemoto2020, Hikosaka2022} e.g. Si, S, O, etc.), but also on the exact mantle composition (\citealt{McDonough1995, Baker1999, Williams2005} e.g. pyrolite, peridotite, chondritic, etc). 
    With mineralogy from from \citep{Wang2018}, we calculate Earth's iron content to be Fe-mf$_E$  $\sim 31$ wt$\%$.

    To study Earth as an exoplanet, we assume small errors in mass-radius data: $M_E = (5.973\pm0.001) \times10^{24}$ \unit{kg} and $R_E = (6371 \pm 6)$ \unit{km}, while ignoring all other information known for Earth (moment of inertia, minor mineral phases, presence of an outer/inner core from seismology, etc). 
    At this accuracy level, for our inverse model, we opt to couple our interior model with a Markov chain Monte Carlo (MCMC) sampler \citep{Foreman2013}. 

\begin{figure}
        \centering
        \includegraphics[width=\linewidth]{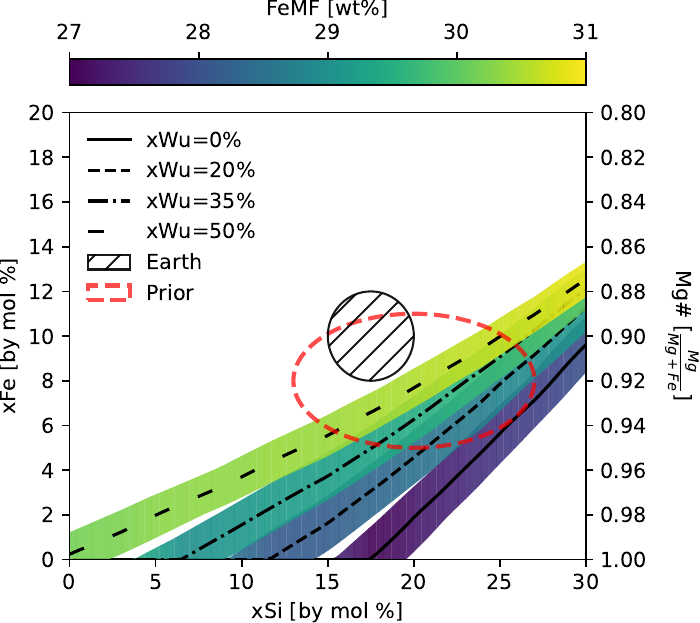}
        \caption{{Relationship between the amount of Si in the core ($\mathrm{xSi}$) and Fe in the mantle ($\mathrm{xFe}=1-\mathrm{Mg}\#$) that fit Earth's mass, radius and cmf ($32.5\pm0.3$ \citep{Wang2018}) for different assumed values of magnesiowustite ($\mathrm{xWu}$) in the lower mantle (colors). The two quantities trade-off effectively as they exchange the iron content from core (left lower corner) to mantle (right upper corner), although total iron budget is not conserved. The black shaded region is the Earth reference for $\mathrm{xSi}$ (\citealt{Fischer2015, Zhang2018, Umemoto2020, Hikosaka2022}) and $\mathrm{Mg}\#$ (\citealt{McDonough1995, Baker1999, Williams2005, Mattern2005, V-C2017, Houser2020}) values, the red region is the best-Earth prior ($\mathrm{xFe} \sim \mathcal{N}(8,3)$, $\mathrm{xSi} \sim \mathcal{N}(20,7)$) we chose to retrieve Earth's cmf.}}
        \label{fig:xsixfe}
    \end{figure}

    The first step, however, is to determine the possible mineralogy values ($\chi$) that reproduce Earth's cmf with our forward model. 
    Specifically, we obtain the plausible values for different degrees of differentiation, represented by xFe= $1 - \mathrm{Mg\#}$; amount of alloys in the core, represented by xSi; and values of Mg/Si in the mantle, captured in $\mathrm{xWu}$ that reproduce the Earth's cmf within one sigma, see Figure \ref{fig:xsixfe}. 
    This exercise informs us of possible priors in mineralogy that allows us to retrieve Earth's values with our simple interior structure model.   
    These $\chi=\left(\mathrm{xfe},\mathrm{xSi}, \mathrm{xWu}\right)$ values are shown in Figure \ref{fig:xsixfe}.
    Unsurprisingly, we find that it is possible to obtain the same cmf, by trading off xSi and xFe, which effectively moves iron between core and mantle. 
    This exchange, however, is not perfect as it does not conserve the total iron budget (see Fig. \ref{fig:xsixfe}). 
    The specifics of this trade-off depend on the amount of Mg to Si in the mantle (xWu).    

    Guided by this forward modeling and Earth's geochemical constraints (see appendix \ref{appendix_A}), we choose our best fit priors to be $\mathrm{xFe} \sim \mathcal{N}(8,3)$, $\mathrm{xSi} \sim \mathcal{N}(20,7)$ and $\mathrm{xWu} \sim \mathcal{N}(25,5)$  (red ellipse in Fig. \ref{fig:xsixfe}). 
    They produce a cmf of $31.3 \pm 2.5$ wt$\%$, which is consistent with Earth's value (within 1$\sigma$).
    We note that this best fit case requires less iron in the mantle compared to the typical value for Earth. 
    Using Earth's nominal value of $\mathrm{Mg}\# \sim 0.9$ ($\mathrm{xFe}\sim10\%$)  \citep{McDonough1995, Baker1999, Williams2005} instead, produces an excess in mantle density of $\Delta \rho \sim 1\%$ compared to \texttt{PREM} \citep{Dziewonski1981} which leads to a lower cmf than Earth's value. 
    This mantle density excess is also found in other studies (\citealt{Mattern2005,V-C2017,Houser2020}).
    {In terms of the iron-mass fraction, our simple mineralogy treatment, that consists of only major chemical phases (Mg, Si, Fe, O), with Ni in chondritic proportions (Fe/Ni$\sim 16$ \citet{McDonough1995}) and Si as a proxy for light alloys in the core, yields a Fe-mf $= 29 \pm 0.6$ wt$\%$. 
    This value is below that of Earth's by 2 points (Fe-mf$_E\sim 31$ wt$\%$).
    Thus, this exercise informs us of the accuracy by which we can reproduce the Earth and shows that the use of a simple mineralogical model that minimizes the number of assumptions outweighs the loss in accuracy.} 
    
{
    For exoplanets, though, there is a lack of information on how to extrapolate the degree of differentiation, exact core light alloy abundance and even the Mg/Si ratio. 
    That is, $\chi$ is unconstrained. 
    To show the effect of this lack of knowledge, we retrieve Earth's parameters by using uninformed priors, $\mathrm{xFe}\#\sim \mathcal{U}(0,20)$, $\mathrm{xSi}\sim \mathcal{U}(0,20)$ and xWu=20$\%$), which we subsequently use for exoplanets (see sec. \ref{results:rocky}).
    These general priors yield a cmf of $27.9 \pm 2.5$ wt$\%$.
    However, this result underestimates the cmf error as the true error should reflect the dispersion arising from the extremes in $\chi$ equally (at $\sim 3\sigma$).
    Meaning, that any values of $\chi$ are equally plausible and one should not favour the most probable cmf that arises from the various combinations of xFe, xSi as well as xWu that lead to the same cmf.
    That is, by considering the end-members in $\chi$,  we obtain the true error in cmf and Fe-mf, which are $\pm 7$ points and $\pm 3-5$ points respectively (see appendix \ref{appendix_B} for details). 
    These errors are the intrinsic floor errors arising from our ignorance in mineralogy of other rocky planets.
}
        
    \begin{figure}
        \centering
        \includegraphics[width=\linewidth]{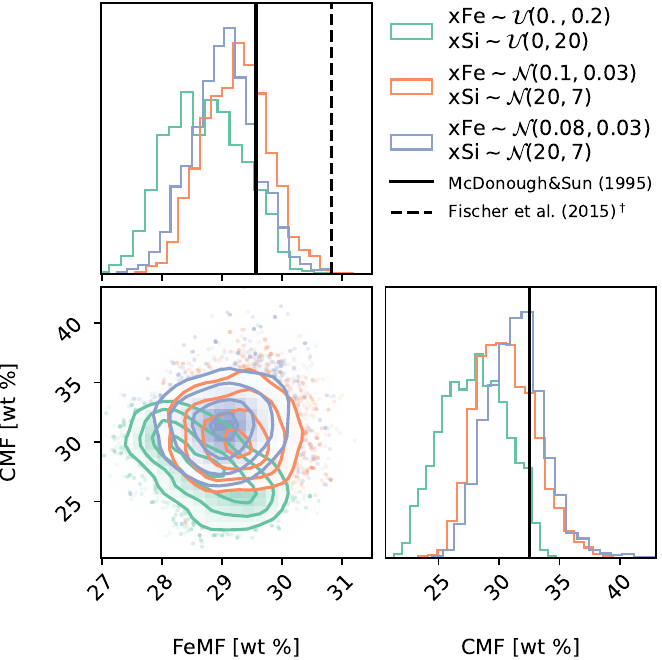}
        \caption{{Earth as an exoplanet: Corner plot of the cmf and Fe-mf posteriors assuming different priors: uniform prior ($\mathrm{xFe} \sim \mathcal{U}(0,20)$, $\mathrm{xSi}\sim \mathcal{U}(0,20)$ and $\mathrm{xWu}=20\%$) in green, Gaussian prior ($\mathrm{xFe} \sim \mathcal{N}(10,3)$, $\mathrm{xSi}\sim \mathcal{N}(20,7)$ and $\mathrm{xWu}\sim \mathcal{N}(20,5)$) in orange, iron-reduced Gaussian prior ($\mathrm{xFe} \sim \mathcal{N}(8,3)$, $\mathrm{xSi}\sim \mathcal{N}(20,7)$ and $\mathrm{xWu}\sim \mathcal{N}(20,5)$) in blue. The vertical lines represent the Earth's expected Fe-mf values of 29.5 (solid black \citet{McDonough1995}) and 30.8 (dash black  \citet{Fischer2015}), while the cmf expected value is 32.5$\pm 0.3$ wt$\%$ \citep{Wang2018}.}}
        \label{fig:earth}
    \end{figure}
    
    {
    We summarize the results from the different assumptions on mineralogical priors in Figure \ref{fig:earth} and we present additional Earth runs in appendix (Table \ref{Table:summary1}). 
    We find that for \emph{extremely precise} mass-radius data, inferences in cmf and Fe-mf depend on the assumed mineralogy. 
    Our exercise of looking at Earth as an exoplanet suggests the intrinsic error is $\sim 7$ wt$\%$ in cmf and $\sim 3-5$ wt$\%$ in Fe-mf due to uncertainties in terrestrial interior composition (degree of differentiation, exact alloy in the core and Mg/Si ratio). 
    This shortcoming applies to all interior structure models and dedicating more observing time to constrain $\sigma_\mathrm{cmf}<7$ or $\sigma_\mathrm{Fe-mf}<5$ may be counterproductive. 
    As a corollary though, there is a mass and radius precision beyond which these assumptions do not matter (see next section).}

\section{Results}
\subsection{Purely rocky exoplanets}
    We first consider purely rocky planets. 
    We choose a 5$M_\oplus$ mass planet as our nominal case and consider different mass and radius errors. 
    We obtain the core-mass fraction, iron-mass fraction, silicate-mass fraction and magnesium mass-fraction using our fiducial rocky model (see Sec. \ref{model:MC}). . 
    We obtain the relationship between uncertainties in mass, radius and composition for Earth-like and Mercury-like planets.
    
    We present the effects of error in cmf (top panel) and Fe-mf (bottom panel) in Figure \ref{fig:rocky}, where the colour bar represents the error in the derived distribution (in weight percent). 
    The error in Fe-mf due to radius and mass uncertainties, as well as the error floor due to modeling is a couple of points lower compared to cmf error.
    
\subsubsection*{Earth-like vs Mercury-like Planets}
\begin{figure*}
        \centering
        \includegraphics[width=\linewidth]{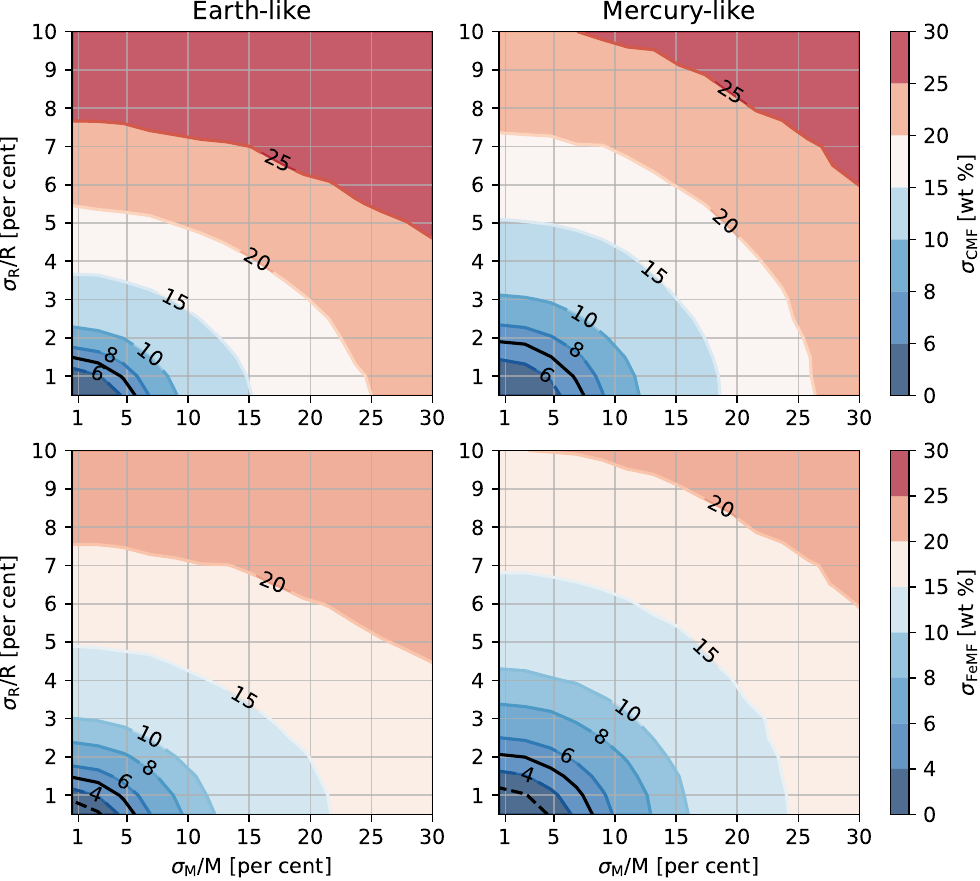}
        \caption{Effect of M-R observation uncertainties on the error in the distribution of cmf (\emph{top}) and Fe-mf (\emph{bottom}) .  \emph{Left}: cases for Earth-like (cmf$=33$ wt$\%$) composition. \emph{Right}: cases for Mercury-like composition (cmf$=74$ wt$\%$). Observing an Earth-like planet will require higher precision to achieve the same error in cmf/Fe-mf compared to Mercury-like planets. The solid black contours represent the threshold below which mineralogy assumptions affect inferences, corresponding to cmf of $\pm7$ and Fe-mf of $\pm5$ points. The dashed black contour is for the Fe-mf of ±3 points case (see Appendix  \ref{appendix_B} for details). Above this threshold, the error in cmf and Fe-mf are mainly set by the errors in M–R.}
        \label{fig:rocky}
    \end{figure*}

\begin{figure}
        \centering
        \includegraphics[width=\linewidth]{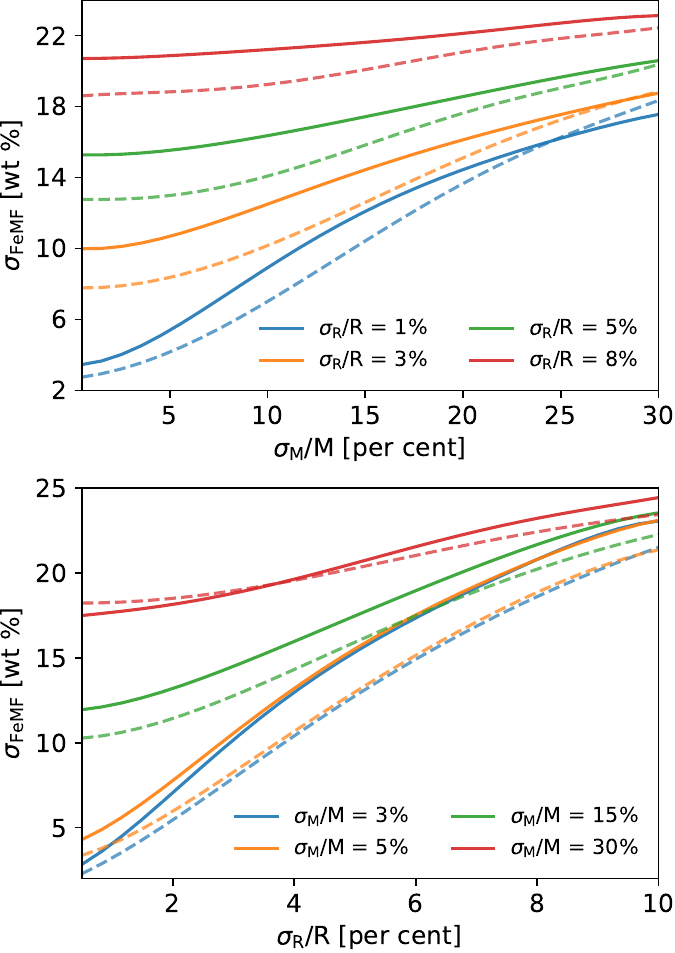}
        \caption{{Effect of M-R observation uncertainties on the error of the Fe-mf distribution given different radius uncertainties (\emph{top}) or mass uncertainties (\emph{bottom}) . The solid lines correspond to Earth-like planets, while dotted lines correspond to Mercury-like planets.}}
        \label{fig:rocky_error}
    \end{figure}

    The contour maps we show in Figure \ref{fig:rocky} can be used to determine the precision in observation required to achieve a target composition precision. 
    For example, to constrain the Fe-mf error to 8 wt$\%$ (10$\%$ in cmf) with $\sigma_R/R=2\%$ the mass uncertainty has to be $5\%$ for an Earth-like planet (cmf$=33\pm10$ wt$\%$) and $11\%$ for a Mercury-like planet (cmf$=74\pm10$ wt$\%$). 
    Observing this example planet for longer to improve the mass will provide minimal gains at too high an observational cost as the radius uncertainty will limit any inferences. 
    {That is, for a given radius uncertainty ($\sigma_R/R$), there is a mass uncertainty ($\sigma_M/M$) below which it is not prudent to keep observing. 
    We demonstrate in Figure \ref{fig:rocky_error} the improvements in Fe-mf error due to changes in mass uncertainty given a fixed radius uncertainty and vice versa. 
    The moment when the uncertainty effect has a gradient close to zero (horizontal segment) indicates the target error in mass or radius for observation.}

\label{results:rocky}

    As previewed in our simple analytical model (sec. \ref{model:simple}) Mercury-like planets require less precise measurements in mass to yield the same compositional error constraint compared to that of Earth-like planets, although the difference is typically small--around (6$\%$ difference in $\sigma_M/M$ values), they may increase (to $10\%$ or more) for planets that have larger radius uncertainty.
    It is important to note that for an error in radius of $\sigma_R/R \gtrsim 5\%$ the improvement in mass error has to be substantial to improve the Fe-mf precision, especially for Earth-like planets.
    On the other hand, when both radius and mass measurements are highly precise, there is little difference in the error of the inferred Fe-mf ($<1$ points) between Earth and Mercury-like planets.

    In a similar fashion, planets that are iron depleted with respect to Earth-like will have a larger $\sigma_{\mathrm{Fe-mf}}$ for the same errors in radius ($\sigma_R/R$) and mass ($\sigma_M/M$).
    These results fall in line with our simple analytical prescription (Sec. \ref{model:simple}), where the planets have two distinct densities: iron (mostly in cores) and silicate oxides (in mantles).
    The Fe-mf error arises primarily due to the difference in densities of both materials, while pressure effects are secondary.
    We further investigate error behaviour with planetary mass, guided by the fact that $\sigma_\mathrm{Fe-mf}$ has an implicit mass dependence in our simple analytical prescription. 
    We test the same two planetary compositions using 3-$M_\oplus$ planets instead and find that the Fe-mf error is almost identical and exhibits similar trends. 
    Therefore, total mass of the planet has little ($<1$ Fe-mf point) or no effect on the inferred Fe-mf and cmf error. 

\subsubsection*{{Effect of Mineralogy Priors}}
\label{results:prior}
{
    We are interested in establishing the conditions when $\chi$ prior matters for exoplanets and the subsequent effect on the interior parameters: Fe-mf, Si-mf, Mg-mf and cmf.
    We find that there is an uncertainty threshold of $\sigma_R/R>2\%$ and $\sigma_M/M>10\%$, above which the mineralogy prior has little effect on the retrieved interior parameters. 
    This arises from the fact that below this threshold the resulting error in cmf and Fe-mf is comparable to the intrinsic floor error (see section \ref{Earth}).
    Thus, for exoplanets that are above the threshold, any model prior is viable and the posterior distributions of cmf or Fe-mf should fall within 1 sigma.
    We indicate this error threshold in Figure \ref{fig:rocky} with black contours, below which careful consideration on mineralogy priors ($\chi$) is required.
    For exoplanets with modest and large errors, this finding justifies using simple models with fixed chemistry-- such as pure iron cores and pure Si-Mg mantle, where Fe-mf = cmf  (e.g. \citealt{Valencia2006, Seager2007, Dorn2015, Santos2015, Zeng2016, Brugger2017}) to constrain exoplanet composition. 
    However, for planets with exquisite data, like those in the TRAPPIST-1 system \citep{Agol2021}, different mineralogical priors will lead to different inferences and should be carefully implemented.
    }
\subsubsection*{{Best interior metrics: Fe-MF $\&$ CMF}}
\label{results:femf}

    There are two ways to directly compare the composition of exoplanets to that of their host star: either in refractory ratio space (e.g. Fe/Mg and Fe/Si) or in interior parameters space (e.g. cmf/Fe-mf). 
    It is unclear which is a better metric for comparison given that mineralogy assumptions ($\chi$), as well as radius and mass uncertainties affect these metrics differently.

    In general terms, a mass-radius measurement constrains the amount of heavy to light material in a planet. 
    The main planetary constituents of heavy material in rocky planets are Ni and Fe, whereas the light materials are Si, Mg, O and additional less abundant elements (not considered in our model). 
    Thus, for a given M-R the amount of Fe+Ni and Mg+Si+O is robustly constrained. 
    To discern the abundance of each element individually, however, one has to invoke assumptions on mineralogy. 
    This explains why retrieving the amount of Si or Mg alone as well as their ratios is subject to mineral assumptions. 
    The same in principle applies to Fe and Ni, but with the added constrain that both elements behave similarly (metallic elements) and Ni is produced less abundantly than Fe, by at least an order of magnitude. 
    Thus, that there is at least partial justification in making assumptions on the Ni/Fe ratio of the planet, which would narrow the total Fe budget.
    In contrast, the cmf is a metric that is set both by density contrasts (heavy to light material) and mineralogy assumptions, owing to the fact that there are light alloys in the core. 
    But since the core is still dominated by iron, the cmf is a metric that follows closely the advantages of Fe-mf. 
    The same is not the case for Si-mg and Mg-mf or their their ratios. 
    This explains why Fe-mf and cmf are comparably influenced by M-R uncertainties, with Fe-mf having somewhat lower errors overall. 
    However, as mentioned, if one assumes a consistency in Ni/Fe ratio between planet and host star, which is standard practice in geochemical models \citet{McDonough1995}, the error in Fe-mf is reduced in half. 

    Moreover, the star's Fe-mf equivalent, which is derived from stellar ratios Fe/Si, Fe/Mg and Mg/Si but not mass or radius, is highly invariant to mineralogy priors compared to the stellar cmf, Mg-mf or Si-mf. 
    These reasons makes Fe-mf the most robust metric to make comparisons between planets and their host stars. 
    On the other hand, one advantage of comparing planets and stars within refractory chemical ratios is that once the planet's composition is constrained the comparison is straightforward as the star's are given and need no assumptions.
    However, refractory ratios may be misleading if the planets are very iron rich or iron poor given the non-linear behaviour of the these ratios and they also have larger overall errors as they are a combination of individual errors in Fe, Mg and Si abundances.

    In conclusion, both cmf and Fe-mf metrics are equally robust for exoplanets, while for stars the Fe-mf equivalent is a better parameter. 
    This suggest that the best metric to compare planets and stars is the iron-mass fraction. 
    To note, for exoplanets with huge mass and radius uncertainty the assumptions in mineralogy are unimportant and it is reasonable to assume that the core and mantle are made purely from Fe and Mg-Si rock, respectively, thus yielding an equivalence between cmf and Fe-mf. 
    For straightforward stellar cmf/Fe-mf calculations we have made our code publicly available: \href{https://github.com/mplotnyko/SuperEarth.py}{https://github.com/mplotnyko/SuperEarth.py}.
    
\subsection{Water Worlds}
\label{results:water}
    Similar to our analysis of rocky planets, we use our sampling algorithm to analyze the effects of observation uncertainty on constraining the amount of H$_2$O (wmf) for temperate or cold planets.     
    We consider three different constraints on the rocky interior: fixed refractory ratios, stellar priors and unconstrained priors to investigate their influence on inferences in the volatile content (wmf).

\subsubsection*{Fixed rocky interior}
    To begin with, we fix the refractory ratios  to $\mathrm{Fe/Mg}=1.7$, $\mathrm{Mg/Si}=0.8$, informed by the mean value in the stellar chemical distribution (from APOGEE catalog \citealt{Majewski2016,Ahumada2020}). 
    This translates to fixing the mantle-to-iron core mass ratio, making the goal of constraining water from mass and radius measurements non-degenerate.

    \begin{figure*}
        \centering
        \includegraphics[width=\linewidth]{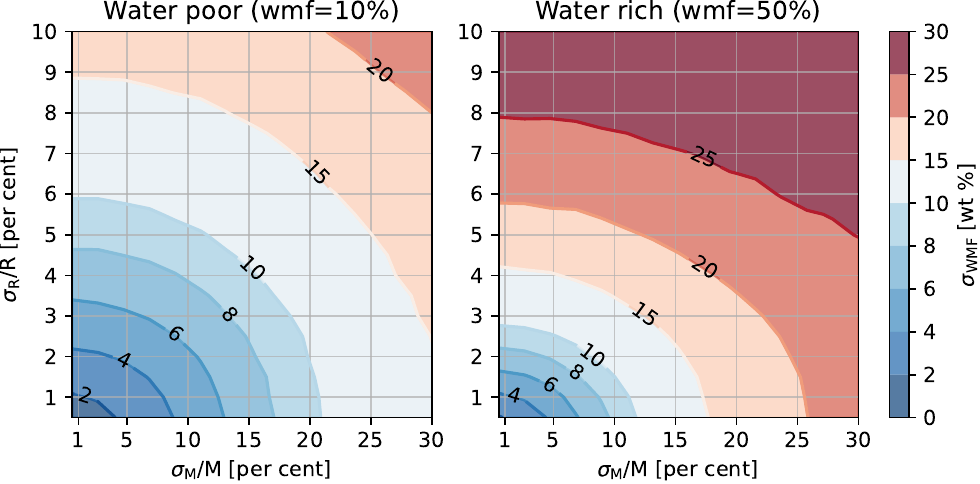}
        \caption{Effect of M-R observation uncertainties on the error in the wmf distribution (colorbar). \emph{Left}: cases for  a water-poor planet (wmf $=10$ wt$\%$). \emph{Right}: cases for a water-rich planet (wmf $=50$ wt$\%$). We assume a fixed rocky interior with $\mathrm{Fe/Mg} = 1.7$, $\mathrm{Mg/Si} = 0.8$. Water-poor targets have more precise water abundance predictions for the same errors in M-R than water-rich planets.}
        \label{fig:water}
    \end{figure*}

\begin{figure}
        \centering
        \includegraphics[width=\linewidth]{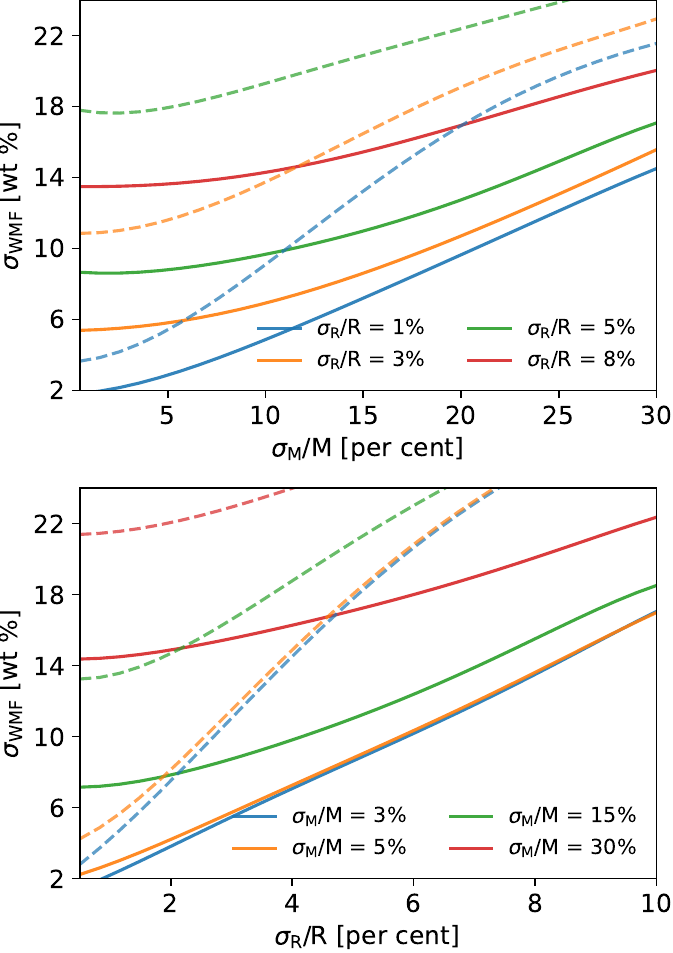}
        \caption{{Effect of M-R observation uncertainties on the error of the wmf distribution given different mass uncertainties (\emph{top}) or radius uncertainties (\emph{bottom}). The solid lines represent the water-poor case, while dotted lines represent the water-rich case.}}
        \label{fig:water_error}
    \end{figure}

    The results are shown in Figure \ref{fig:water} for the two planetary cases considered: water-poor with 10 wt$\%$ wmf and water-rich with 50 wt$\%$ wmf and can be used to guide observational campaigns as to achieve a desired compositional accuracy for cold and temperate water worlds.{ Another way to see these results is shown in, Figure \ref{fig:water_error} where it is apparent how improvements in wmf error are subjected to improvements radius or mass measurements.} 
    We find that errors for water-poor planets are consistently lower than for water-rich planets. 
    The latter can have $\sigma_\mathrm{wmf}$ up to 10 wt$\%$ higher wmf error for the same observation uncertainty.
    
    From Figure \ref{fig:water} for example, to obtain a desired wmf error of $\sigma_\mathrm{wmf}=10$ wt$\%$, a planet with $\sigma_R/R=2\%$ will require an error mass error of only $\sigma_M/M=20\%$ for a water poor-planet versus an error mass of  $\sigma_M/M=9\%$ for a water-rich planet.
    Namely, planets with higher uncompressed densities (density when pressure effects are ignored), such as those with $\mathrm{wmf} = 10$ wt$\%$ or super-Mercuries, have better compositional estimates when compared to lighter (smaller bulk density) planets.
    This effect can be seen with the simple analytical prescription for water-worlds we derived (see Sec. \ref{model:simple}), although it underestimates $\sigma_\mathrm{wmf}$ and is not as accurate as the one for rocky planets.

\subsubsection*{Constraints from Stellar Priors}
    In the second case, we use the stellar \emph{distribution} values as a prior on the \emph{distribution} of the rocky interior. 
    We use the Fe/Mg stellar population values from APOGEE catalog (\citealt{Majewski2016,Ahumada2020}) and adopt a $\mathrm{Fe/Mg} \sim \mathcal{N}(1.7,0.4)$ distribution, emulating a prior in the refractory composition should the host star be observed.
    The difference between the fixed and constrained case is negligible ($\sim 1$ wt$\%$) for most uncertainty values, except for small mass-radius errors where the differences are somewhat noticeable.
    We find that if $\sigma_R/R<2\%$ and $\sigma_M/M<10\%$, $\sigma_\mathrm{wmf}$ is larger by 1 to 4 wt$\%$ when compared to the fixed case.
    We also looked at having a less precise prior on the Fe/Mg ratio ($\mathrm{Fe/Mg} \sim \mathcal{N}(1.7,0.7)$) and find that there is barely any noticeable change in wmf error ($<0.5$ wt$\%$ \ref{fig:AllWMF}). 
    This means that refractory ratios errors that are of the same magnitude as the Sun's errors (up to 2 times) are acceptable and more precise stellar values are not required to constrain planet water content.

\subsubsection*{Unconstrained prior}
    Lastly, we consider the completely unconstrained case where cmf and wmf can adopt any values.
    This problem is known to be degenerate \citep{Valencia2007}, but our interest here is to determine the minimum in error that can be achieved with only mass and radius information.
    We randomly draw some cmf value from a uniform distribution for every sampled mass-radius pair and subsequently solve for the wmf that fits the mass and radius.
    We find that the degeneracy is so strong that it is not possible to constrain the $\sigma_\mathrm{wmf} \sim 15$ wt$\%$, regardless of the precision in mass and radius. 
    Therefore, the only possible way to constrain water content is to know something about the rocky interior. 
    It is imperative then to determine whether or not the formation of the rocky components of planets, including volatile planets, are correlated to the disk primordial composition. 
    If we assume this to be true and planets have similar refractory ratios to that of their host stars, then even measurements with large uncertainties will constraint the water content. 
    However, the limiting factor in water abundance retrieval for lower mass planets  ($\sigma_R/R=4\%$ and $\sigma_M/M=15\%$) at the moment is not the observation precision of mass and radius, but the absence of refractory measurements of stars.

\section{Discussion}
    In this study we have considered both rocky and water planets, here we discuss implications in our analysis for both cases, due to model/design assumptions. 
        \paragraph*{Rocky Planets:} 
        While assumptions in mineralogy do not affect planets with moderate to large error bars, they are important for planets with very precise data. Knowing the boundary is important and it depends on the actual bounds to possible mineralogies. For rocky planets we established that three parameters matter: the degree of differentiation, core alloy and Mg/Si mantle ratio. For the degree of differentiation the pure iron core is an obvious maximum, but the minimum is less clear. The lowest $\mathrm{Mg}\#$ the Earth may have corresponds to a core-less planet with all iron mixed in with oxides with a value of $\mathrm{Mg}\#=0.55$. However, studies on planetesimals have shown that depending on conditions for growth and cooling, planetesimals may be less differentiated with more iron in the mantle ($\mathrm{Mg}\#<0.8$ \citet{Sramek2012}). It is unclear how growth from planetesimals to planets alter the $\mathrm{Mg}\#$ but it can be expected that it increases due to the effect of heating during impacts. For the problem of core alloys, it is uncertain if there is an upper limit to the amount of light elements that can be in the core ($\mathrm{x_{light}}$) and which ones. Within the terrestrial planets, Earth and Venus have $\mathrm{x_{light}}<20\%$ by mol while Mars and Mercury appear to have $\mathrm{x_{light}}>20\%$  (\citealt{Mars_2021,Berrada2022}). {For the case of mantle Mg/Si, we need to establish if the Mg/Si ratio of the planet follows that of the star and if so, what minerals are present for stars with Mg/Si <1. The former is not an easy task, but perhaps from studying solar system terrestrial planets and asteroids, we can better determine if a primordial bulk composition in these elements is a reasonable expectation. Better establishing the bounds in mineralogy for super-Earths with theoretical or laboratory predictions, would feed directly into better inferences on the individual planetary compositions relevant for exquisite mass-radius measurements expected in the next decades.}    
        \paragraph*{Water Worlds:} To infer their water content it is imperative to have information about their rocky interior. In this work we show that without this information, the most optimistic constrain would be 15 wt$\%$ for wmf. In the absence of observations, the most straight forward assumption for the rocky interior has been to use the refractory ratios of the star, but this link has to be further investigated and confirmed.
        \paragraph*{Vapour Worlds:} In this study we were focused on habitable or cold planets specifically (water is in liquid/ice state), while ignoring the possibility of warm/hot water worlds. Due to its effect on radius \citep{Turbet2020}, the presence of vapor can alter water content estimate by orders of magnitude. For example, a planet that has $M=5M_E$ and $R=1.6 R_E$, will have wmf$=10$ wt$\%$ for condensed water case and wmf$=1$ wt$\%$ for gaseous case. Our results can be seen as an upper limit for constraining water on vapour planets. We leave for future work the investigation of the relation for vapor and hycean worlds. 

\section{Conclusion}
    This is a summary of our key findings:
    \begin{itemize}
        \item {We obtain the relationship between error in mass, error in radius and inferred error in Fe-mf  for rocky planets and wmf for water worlds. Our results can guide observational efforts to meet target goals in Fe-mf/wmf errors, as well as to determine when more observing time is unnecessary. We provide all the necessary codes in \href{https://github.com/mplotnyko/exopie}{https://github.com/mplotnyko/exopie}. 
        \item Interior structure models for rocky planets have an intrinsic error when estimating composition arising from mineralogy uncertainties such as the exact alloy in the core, the degree of differentiation and the mantle or overall Mg/Si ratio. This intrinsic error is around 5 wt$\%$ in Fe-mf and 7 wt$\%$ in cmf.}
        \item  {Due to this intrinsic error from mineralogy, to study exoplanets with exquisite data ($\sigma_R/R<2\%$ and $\sigma_M/M<10\%$) it is important to carefully assess the mineralogical assumptions. Planets with moderate to large errors, which are the majority of exoplanets, have Fe-mf's that are 2 to 3 times larger than this intrinsic error such that details in the mineralogy do not matter.} 
        \item Denser planets are easier to constrain, while lighter planets are more difficult. That is, for the same mass and radius certainties, Earth-like planets have slightly higher Fe-mf error ($\sim 2$ wt$\%$) compared to Mercury-like, while wmf error is almost twice for water-rich planets compared to water-poor planets. 
        \item {Planetary mass and radius measurements constraint the heavy to light material. For rocky planets that is the Fe+Ni to the Mg+Si+O abundances, while constraints on individual elements or their ratios, depend on chemical modeling assumptions. However, due to similarities in behaviour of Ni-Fe and the overabundance of Fe with respect to Ni, mass and radius measurements robustly constrain the Fe-mf.}
        \item {In addition, translating the refractory ratios of the star to an Fe-mf equivalent is invariant to mineralogical assumptions. Thus, it is better to make stellar-planet composition comparisons in Fe-mf space. }  
        \item Constraining water content (wmf) for water worlds requires insight about planets rocky interior, since otherwise (unconstrained case) wmf error cannot be less than 15 wt$\%$. Should stellar refractory ratios be used, current precision levels in stars are enough to constraint the water content. However, this highlights the need to firmly establish any links between stellar and planetary refractory compositions. 
    \end{itemize}    

\section*{Acknowledgements}
    We would like to thank Dr. Angie Wolfgang for the useful discussions about statistical methods and the anonymous reviewer for their comments that have helped improved our study. This work has been partially funded by the Natural Sciences and Engineering Research Council of Canada (Grant RGPIN-2021-02706). We would like to acknowledge that our work was performed on land traditionally inhabited by the Wendat, the Anishnaabeg, Haudenosaunee, Metis and the Mississaugas of the New Credit First Nation. 
    
\section*{Data availability}
    The data underlying this article will be shared on reasonable request to the corresponding author.
    While, the corresponding code is available at \href{https://github.com/mplotnyko/exopie}{https://github.com/mplotnyko/exopie}.

%%%%%%%%%%%%%%%%%%%%%%%%%%%%%%%%%%%%%%%%%%%%%%%%%%

%%%%%%%%%%%%%%%%%%%% REFERENCES %%%%%%%%%%%%%%%%%%

\bibliographystyle{mnras}
\bibliography{bibliography} % bibtex with file bibliography.bib

%%%%%%%%%%%%%%%%%%%%%%%%%%%%%%%%%%%%%%%%%%%%%%%%%%

%%%%%%%%%%%%%%%%% APPENDICES %%%%%%%%%%%%%%%%%%%%%

\appendix
\section{Exoplanet Mineralogy}
\label{appendix_A}

    For rocky interiors, we consider a two layer model with the core at the center and a mantle on top. 
    Specifically, we assume that the core is made of a Fe-Ni-Si alloy and is assumed to be solid, while the mantle is divided into upper, lower mantle and lowermost mantle.  
    For the core we assume that Si can range between $0-20\%$ by mol, thus representing the pure end-member case of Fe-Ni core and a lighter core similar to Mars or Earth (xSi$=20\%$ by mol \citealt{Khan2018,Fischer2015}).
    For the mantle layers, we assume that the upper mantle is made of olivine (wadsleyite/ringwoodite at higher pressures) and pyroxenes, the lower mantle of bridgemanite and magnesiowustite (or wustite for short) and the lowermost mantle of post-perovskite and magnesiowustite.
    In addition, for these mantle minerals, we consider both magnesium and iron phases in proportion to the global magnesium number $\mathrm{Mg}\#=\frac{\mathrm{Mg}}{\mathrm{Fe}+\mathrm{Mg}}=1-\mathrm{xFe}$.
    For Earth, geochemical constraints place a range of $\mathrm{Mg}\# = 0.88-0.92$ \citep{McDonough1995, Baker1999, Williams2005}, in the mantle and xSi=$15-20\%$ by mol in the core \citep{Fischer2015, Zhang2018, Umemoto2020, Hikosaka2022}. 
    For exoplanets, instead we consider a wider range of compositions. 
    Namely $\mathrm{xFe}=0-20$ by mol based on a completely differentiated planet ($\mathrm{Mg\#}=1$) and one differentiated in a similar degree to that of Mars ($\mathrm{Mg}\# \sim 0.75$ \citep{Khan2018}).

    We ignore all the Ca-Al mantle phases as they occupy a small fraction of mass ($0-2\%$) and do not affect the interior structure as seen in the density profile, but may alter the total iron budget due to the replacement of Fe bearing rocks.
    We vary the content of wustite when looking at the Earth as an exoplanet to understand its effect, but assume a value of $20\%$ of wustite by mol (xWu) for both lower and lowermost mantle for exoplanets. This fixes the mantle Mg/Si ratio to 0.86. 
    This is in-line with Earth's xWu values ranging between 0-35$\%$ \citep{McDonough1995, Baker1999, Williams2005} and the most accepted value of $\sim20\%$ \citep{Hirose2006, Stixrude2011}.
    It is important to note that we do not account for mineralogy composition that produce Mg/Si $<0.7$ values, due to the lack of EOS data at relevant pressures and temperatures.

\section{Effects of Mineralogy}
\label{appendix_B}
    Here we expand on how mineralogy effects the interior composition metrics of planets, such as the cmf, Fe-mf, Si-mf and Mg-mf. 
    To isolate the effects of mineralogy priors on each of this metrics, we fix the mass and radius to that of Earth's and vary each component of the chemical vector  $\chi =\left( \mathrm{xFe}, \mathrm{xSi}, \mathrm{xWu}\right)$.    
    We find that: 
    \begin{enumerate}[1)]
        \item For a given $\mathrm{xWu}$ value, the Fe-mf is almost invariant to assumptions on $\mathrm{xFe}$ and $\mathrm{xSi}$ ($\Delta$ Fe-mf$ \sim 1$ wt$\%$), while affecting more strongly the cmf prediction ($\Delta$ cmf $ \sim 6$ wt$\%$) . 
        \item Considering different $\mathrm{xWu}$ values have an effect on Fe-mf ($\Delta$ Fe-mf$ \sim 2$ wt$\%$) but a negligible effect on cmf. 
        \item Exoplanets may have Mg/Si < 1, which should not affect the cmf but affect the Fe-mf. We estimate that the effect is comparable to that within Mg/Si $=1-2$, increasing the Fe-mf error to $\Delta$ Fe-mf$ \sim 2$  wt\%.
        \item Adding Ca, Al and other phases have an impact on the final mineralogy budget (Fe-mf, Si-mf, Mg-mf), but not on the interior structure (cmf). Thus, assuming a constant cmf we find that adding Ca-Al species in proportion to Earth's abundances decreases Fe-mf by $\sim 1$ wt $\%$ and changing Ni from 0 to 10$\%$ in the core changes the Fe-mf by $\sim 2$ wt $\%$. It is also unclear what these values are for exoplanets without making assumptions on how they are connected to their host star composition. 
    \end{enumerate}

\begin{figure*}
        \centering
        \includegraphics[width=\linewidth]{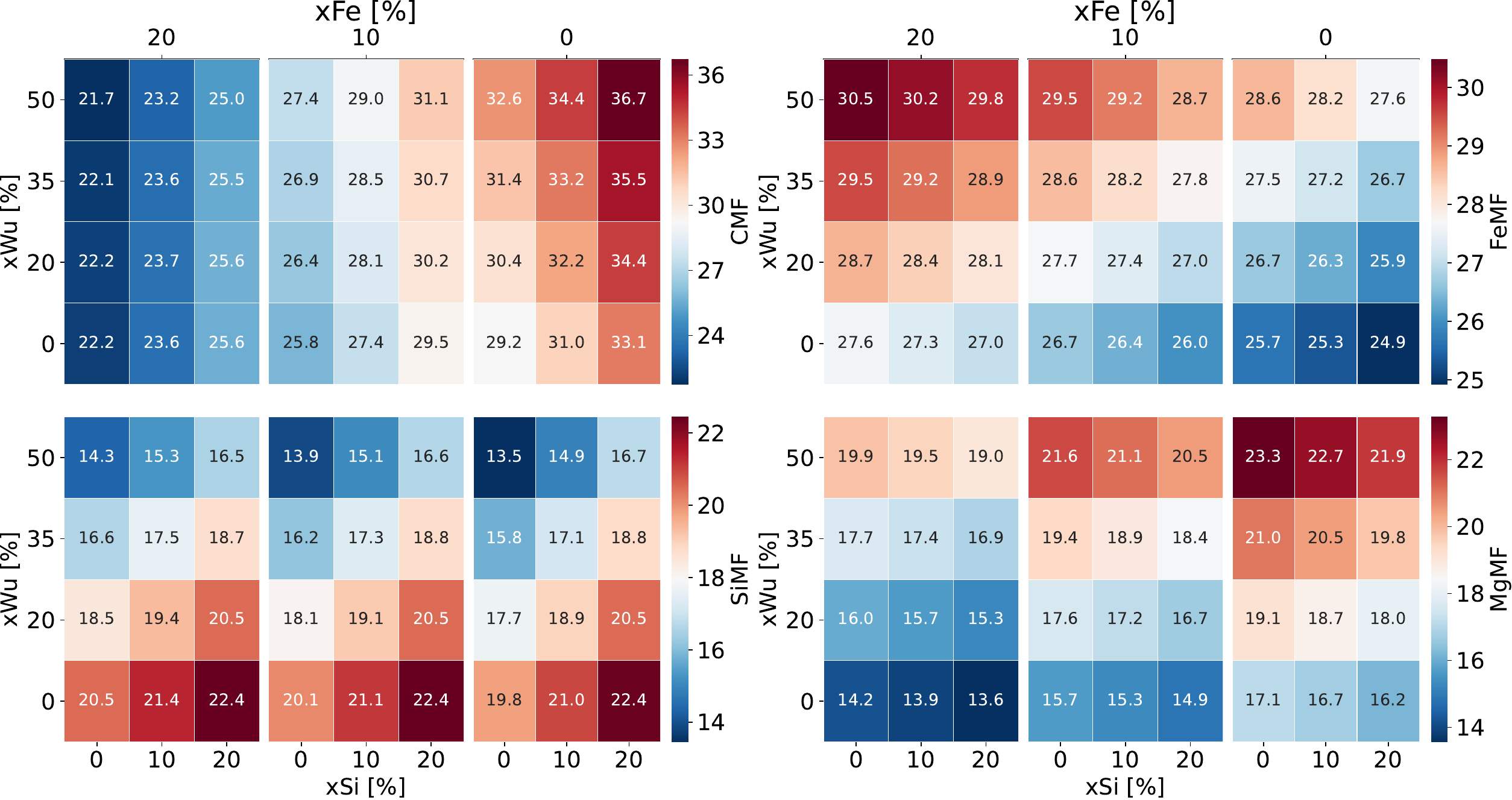}
        \caption{{Effects of mineralogy ($\left( \mathrm{xFe}, \mathrm{xSi}, \mathrm{xWu}\right)$) on the inference of 
        cmf, Fe-mf, Si-mf and Mg-mf. The spread in each parameter is indicated in text and color coded. Notice each parameter has its own colorbar. We split the heatmap into three sections corresponding to xFe $= 20, 10, 0 \%$ by mol (Mg$\#=$ 0.8, 0.9 and 1), which is further subdivided by xSi values of 0, 10 and 20 $\%$ by mol. The spread due to amount of Ni is not accounted in this figure (xNi=$10\%$ by mol). The maximum spread in Fe-mf is $\sim \pm 5$ points, whereas in cmf is $\sim \pm7$ points. }}
        \label{fig:DeltaRes}
    \end{figure*}

    Thus, assuming all the possible $\chi$ we find that the cmf has an intrinsic error of $\pm 7$ points (cmf=22-37), while Fe-mf has $\pm 3-5$ points and Si-mf, Mg-mf have slightly larger error of $\pm 4-6$ points, see Figure \ref{fig:DeltaRes}.
    However, the error in Fe-mf can be lowered by having a prior on the core Ni/Fe or mantle Mg/Si ratio, for example by fixing it to the host star's value. 
    In such a case the intrinsic error in Fe-mf is lowered to $\sim 3$ wt$\%$ 
    An important corollary to this exercise is that interior errors arising from not knowing the planets detailed mineralogy are only important for data in mass and radius that can constrain cmf and Fe-mf better than 7 and 5 points respectively.

    Another parameter, that can be seen as a mineralogy prior is the presence of a crust. For this study we ignored the crust ( with density of $\sim2800\unit{kg/m^3}$ from \texttt{PREM}), and thus overestimated the density of the upper tens of km by $\sim500\unit{kg/m^3}$. Looking at Earth, we determined this has a minor effect on cmf inference of $\Delta\mathrm{cmf}\sim1$ wt$\%$ for a 25 km crust as suggested by \texttt{PREM}. It is unclear how the crustal thickness would vary for other planets. Crustal growth and subsequent thickness are determined by initial formation conditions, such as the thermal state of the mantle or the initial water content. As the planet cools and the resurfaced mantle material crystallizes, the final crust thickness can be anywhere between 5-100 km as seen within the rocky bodies in the solar system \citep{Palin_crust2020}. This problem appears to be less significant for massive super-Earths, given the fact that crustal thickness decreases with planetary mass if all other factors are kept equal \citep{Valencia_crust2009}. The reason is that the ascending mantle material crosses the solidus at shallower depths in massive planets owing to their larger gravities, reducing the melt column and thus, crustal production. Conversely, sub-Earth planets can have overall thicker crusts \citep{Batra_crust2022} and thicker relative crustal sizes. Thus, caution must be exercised when using interior models to infer cmf of small planets as it can change by several per cent. In addition, the amount of water within the mantle and temperature regime are important factors that can change melt production, which are difficult to extrapolate to exoplanets.

\section{}

\begin{figure*}
        \centering
        \includegraphics[width=\linewidth]{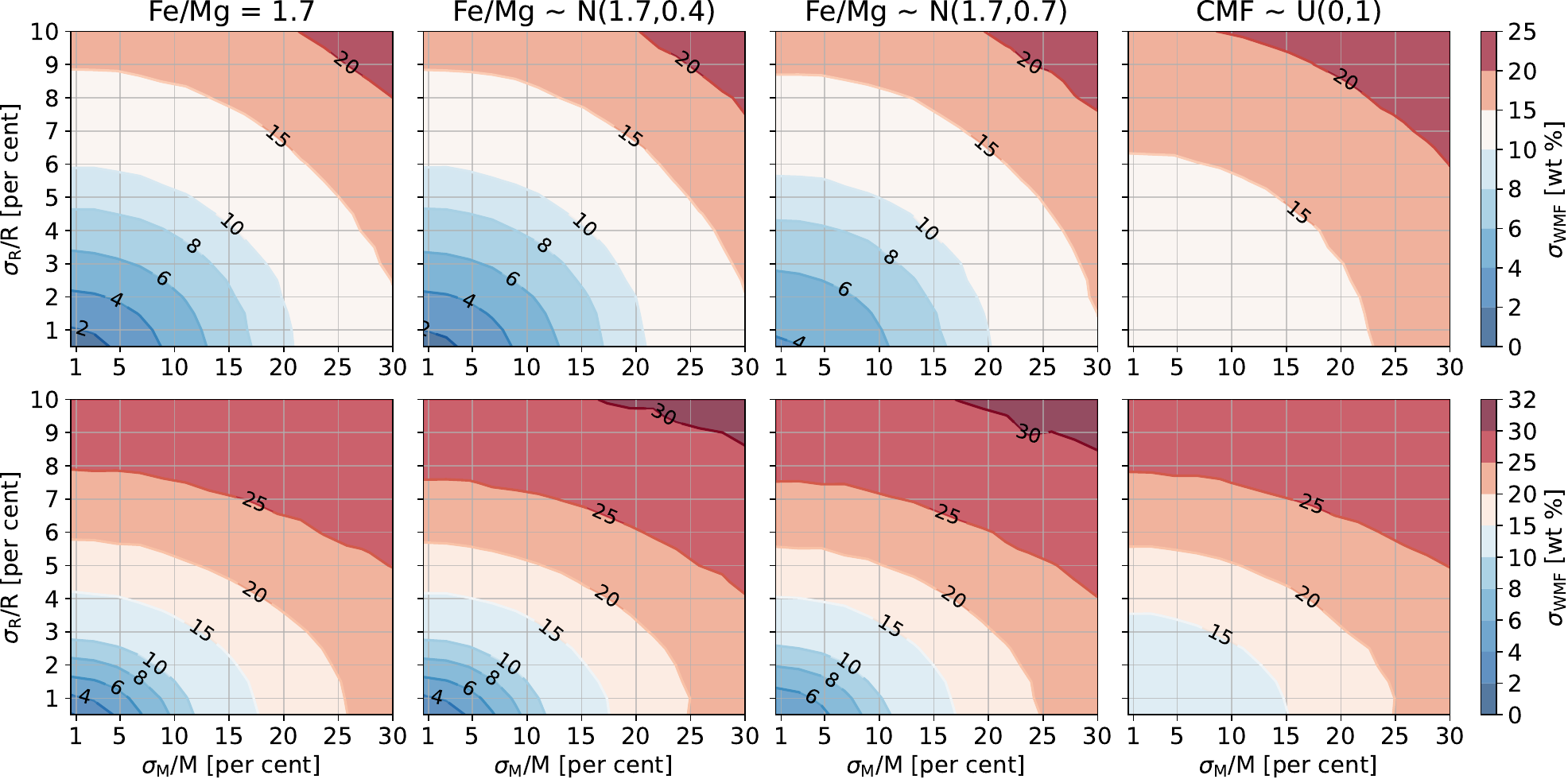}
        \caption{Summary of error effects for water abundant planets: water-poor (top panel) and water-rich (bottom panel). We use different priors for the rocky assumptions: fixed to $\mathrm{Fe/Mg}=1.7$ ($1^{\mathrm{st}}$ column), similar to stars with typical stellar error $\mathrm{Fe/Mg} \sim \mathcal{N}(1.7,0.4)$ ($2^{\mathrm{nd}}$ column), similar to stars with an error that spans all stars $\mathrm{Fe/Mg} \sim \mathcal{N}(1.7,0.7)$ ($3^{\mathrm{rd}}$ column) and completely unconstrained cmf $\sim \mathcal{U}(0,1)$ ($4^{\mathrm{th}}$ columns). We report wmf error for all cases, note the colorbar for top and bottom panel has different scale. The minor differences between the different stellar prior assumptions, show that stars with chemical measurements are accurate enough to constraint the wmf. However, without any priors, the wmf is highly unconstrained. }
        \label{fig:AllWMF}
    \end{figure*}

We present results for Earth as an exoplanets tests in table \ref{Table:summary1}. We report the cmf, Fe-mf and Fe/Si values as well as the Gaussian error we get for these parameters. Additionally, we present summary of error effects for water abundant planets (Fig. \ref{fig:AllWMF}).

\begin{table*}
\caption{Summary of the interior parameters for Earth as an exoplanets tests.}
\label{Table:summary1}
\begin{tabular}{lcccccccccl}
\toprule

Run & xFe $(\%)$ & xSi $(\%)$ & xWu $(\%)$ & CMF (wt$\%$) & CMF Error & FeMF (wt$\%$) & FeMF Error & Fe/Si (w) & Fe/Si  Error \\
UF1  & $\mathcal{U}(0,20)$ & $\mathcal{U}(0,10)$ & 20 &${26.5}$  & $\pm{2.3}$ & 28.8 & $\pm{0.6}$ & ${1.54}$ & $\pm{0.04}$ \\
UF2  & $\mathcal{U}(0,20)$ & $\mathcal{U}(0,20)$ & 20 & ${27.9}$ & $\pm{2.5}$ & 28.7 & $\pm{0.6}$ & ${1.49}$ & $\pm{0.06}$ \\
NF1 & $\mathcal{N}(10,3)$ & $\mathcal{N}(20,7)$ & 20 & ${30.0}$  & $\pm{2.4}$ & 28.7 & $\pm{0.5}$ & ${1.40}$ & $\pm{0.09}$ \\
NV1 & $\mathcal{N}(10,3)$ & $\mathcal{N}(20,7)$ & $\mathcal{N}(25,5)$ & ${30.5}$  & $\pm{2.7}$ & 29.2  & $\pm{0.6}$ & ${1.45}$ & $\pm{0.11}$ \\
NV2 & $\mathcal{N}(8,3)$ & $\mathcal{N}(20,7)$ & $\mathcal{N}(25,5)$ & ${31.3}$  & $\pm{2.5}$ & 29.0 & $\pm{0.6}$ & ${1.44}$ & $\pm{0.11}$ \\

\bottomrule
\end{tabular}
% \begin{flushleft}
% {$\dagger - $ the cmf values follows \citet{Wang2018}}
% \end{flushleft}
\end{table*}

% Don't change these lines
\bsp	% typesetting comment
\label{lastpage}
\end{document}